\documentclass[journal]{IEEEtran}
\usepackage{amsmath,amsfonts,amssymb,mathrsfs}
\usepackage{algorithmic}
\usepackage{algorithm}
\usepackage{array}
\usepackage{textcomp}
\usepackage{stfloats}
\usepackage{url}
\usepackage{verbatim}
\usepackage{graphicx}
\usepackage{cite}
\usepackage{booktabs}
\usepackage{caption}
\usepackage{subfigure}
\usepackage{nomencl}
\usepackage{multirow}
\usepackage{threeparttable}
\usepackage{tabularx}
\usepackage{color}
\usepackage{courier}
\usepackage{pifont}
\usepackage{etoolbox}
\allowdisplaybreaks
\makeatletter
\newcommand{\removelatexerror}{\let\@latex@error\@gobble}
\patchcmd{\@maketitle}
  {\addvspace{0.5\baselineskip}\egroup}
  {\addvspace{-1\baselineskip}\egroup}
  {}
  {}
\makeatother
\removelatexerror

\begin{document}

\title{\huge Quantifying and Optimizing the Time-Coupled Flexibilities at the Distribution-Level for TSO-DSO Coordination}
\author{Yilin~Wen,~\IEEEmembership{Graduate Student Member,~IEEE,}
        Yi~Guo,~\IEEEmembership{Member,~IEEE,}
        Zechun~Hu,~\IEEEmembership{Senior Member,~IEEE,}
        and Gabriela~Hug,~\IEEEmembership{Senior Member,~IEEE}
}

\maketitle

\begin{abstract}
  The flexibilities provided by the distributed energy resources (DERs) in distribution systems enable the coordination of transmission system operator (TSO) and distribution system operators (DSOs). At the distribution level, the flexibilities should be optimized for participation in the transmission system operation. This paper first proposes a flexibility quantification method that quantifies the costs of providing flexibilities and their values to the DSO in the TSO-DSO coordination. Compared with traditional power-range-based quantification approaches that are mainly suitable for generators, the proposed method can directly capture the time-coupling characteristics of DERs' individual and aggregated flexibility regions. Based on the quantification method, we further propose a DSO optimization model to activate the flexibilities from DER aggregators in the distribution system for energy arbitrage and ancillary services provision in the transmission system, along with a revenue allocation strategy that ensures a non-profit DSO. Numerical tests on the IEEE test system verify the proposed methods.
\end{abstract}
\vspace{-10pt}
\section{Introduction}\label{sect:intro}
\IEEEPARstart{E}{lectricity} supply and demand in traditional power systems have been balanced by adjusting the outputs of generators and centralized energy storage systems to track load variations. However, in response to the global energy crisis, power systems worldwide are transforming to integrate more and more renewable energies, such as wind power and photovoltaic (PV), which are highly volatile and thereby increase the demand for balancing services. In this context, power system operators have started to pay more attention to the flexibility offered by distributed energy resources (DERs) on the demand side, e.g., electric vehicles (EVs), battery energy storage systems (BESS), and heat pumps (HPs). Characterized by their flexible operational capabilities, DERs possess the potential to offer substantial flexibility to power systems and to mitigate the variability from renewables. Consequently, activating the flexibility of demand-side DERs in power system operations has become a widespread goal \cite{IEA2019PowerSystem}. 

Due to the small individual capacity and large population, DERs need to be aggregated to a certain capacity for better management and computation from the power system's perspective, leading to the emergence of third-party aggregators who serve as intermediaries between DER users and the power system operators-the transmission system operator (TSO) or the distribution system operator (DSO) \cite{okur2021aggregator}. Aggregating a large number of DERs can provide flexibility for the transmission system operation, such as participating in energy arbitrage and providing ancillary services. However, since DERs are physically located within distribution systems,  adjustments in the power of numerous DERs can affect the distribution system's security, power quality, and operational efficiency. Thus, the DSO also needs to be involved in aggregating the DERs' flexibilities to participate in transmission system operations, i.e., TSO-DSO coordination.

\begin{figure}[!t]
  \centering
  \includegraphics[width=3in]{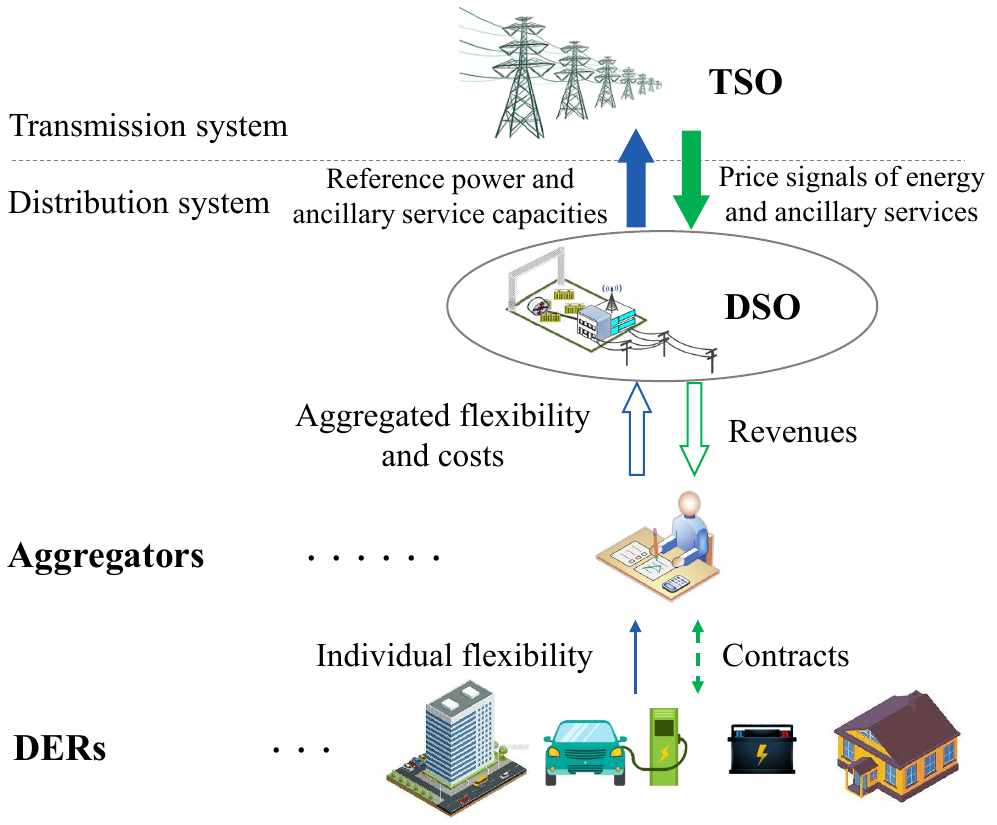}
  \caption{Framework of activating the flexibility of DERs to participate in the transmission system operation \cite{mousavi2021dso,zhang2022optimal,givisiez2020review,talaeizadeh2023prioritization}.}
  \label{fig:framework}
  \vspace{-20pt}
\end{figure}

This paper considers the TSO-DSO coordination framework as depicted in Fig. \ref{fig:framework}. In this framework, the flexibility of individual DERs is first collected by the aggregators, who compensate the DER owners for activating their flexibility through contracts. Aggregators then calculate their aggregated flexibility and its associated costs to report to the DSO. The DSO coordinates the aggregators' flexibility in calculating the total power profile and ancillary service capacities based on the price signals from the TSO. The obtained revenues from the TSO are finally allocated to aggregators. This framework, or similar ones, can be found in many existing studies \cite{mousavi2021dso,zhang2022optimal,givisiez2020review,talaeizadeh2023prioritization}. Ensuring that the DSO is non-profit in this process is essential for the framework to attract DER aggregators to participate and operate sustainably. 

In another TSO-DSO coordination framework, the DSO calculates security restrictions for each aggregator, and aggregators directly participate in the transmission system operation under these restrictions \cite{nazir2021grid,ziras2019mid}. This framework, however, leads to a potential conservativeness of each aggregator's flexibility due to insufficient coordination among aggregators. Therefore, this paper focuses on the framework shown in Fig. \ref{fig:framework} that allows the DSO to coordinate aggregators to provide a larger total flexibility to the transmission system. Within this framework, the primary issue is how to model the aggregated flexibility and quantify its cost and value for the system. In subsequent text, unless specified, the term ``quantification'' refers to quantifying both the cost of flexibility and its value for the system.

Numerous studies have focused on the modeling of aggregated flexibility, proposing models such as the power boundary model \cite{chen2019aggregate}, power-energy boundary model \cite{xu2016hierarchical,wang2022evaluation}, vertex-based polytope \cite{tan2020estimating}, ellipsoid \cite{cui2021network}, generalized battery model \cite{tan2022optimal}, and homothetic polytope \cite{jian2023analytical}. Our previous works \cite{wen2022aggregate,wen2023improved} also proposed efficient modeling methods for flexibility aggregation. Nevertheless, these studies did not discuss how to quantify the costs or values of the feasible regions determined by these aggregated flexibility models, which is important for the DSO's optimal activation of flexibilities. 



Within the field of DSO flexibility markets, quantifying the costs and values of flexibility is always a fundamental part. However, existing studies did not discuss this aspect in detail; instead, they have essentially adopted quantification methods based on the power adjustment ranges in each time slot\cite{morstyn2018designing, mousavi2021dso, zhang2022optimal, tsaousoglou2021mechanism, jian2022dlmp}. For example, reference \cite{morstyn2018designing} proposes to discretize the power adjustment range in each time slot into standardized contract capacities and to define different prices for each segment. References \cite{mousavi2021dso} and \cite{zhang2022optimal} incorporate reserve capacities within the time-coupled flexibility model of aggregators and define their flexibility costs as reserve costs. This quantification method is still essentially based on the power adjustment ranges in each time slot because defining reserves implicitly calculates an approximated power range model (time-decoupled) of the original flexibility model. 

The advantages of using the power adjustment range in each time slot for flexibility quantification are its intuitive calculation and uniformity with generators. However, this method is insufficient for the quantification of the time-coupled flexibility offered by storage-like DERs, e.g., BESSs, EVs, and HPs. In \cite{ziras2019mid}, the opportunity cost is used to quantify the cost of EVs' mid-term flexibility, but this method cannot be directly generalized to various heterogeneous DERs, limiting its application in a unified flexibility activation framework.


In summary, there is a lack of discussion on the quantification method of the cost and value of the time-coupled flexibility model. Filling this gap enhances the willingness of DER users to offer their flexibility to contribute to power system operations. Therefore, this paper proposes a unified quantification method aligned with the time-coupled flexibility models. The term ``unified" here refers to the fact that the proposed quantification method maintains a unified form for the individual and aggregated flexibility of typical DERs. Moreover, we design a DSO optimization model to activate the flexibility of DERs to participate in the transmission system operation for energy arbitrage and ancillary services provision, along with a revenue allocation strategy. The contributions of this paper are fourfold:

\begin{itemize}
  \item From the perspective of DER users, we propose to uniformly compute the flexibility cost based on activated adjustment ranges in both power and accumulated energy consumption. This method is in line with the time-coupled power-energy boundary models of individual DERs' flexibility, which is more suitable for storage-like DERs such as EVs, BESSs, and HPs compared to the traditional power-range-based approach.
  \item From the aggregators' standpoint, we propose to compute their flexibility costs directly according to the activated area of the aggregated flexibility model. Unlike the traditional power-range-based formulations, the proposed approach aligns with the aggregated flexibility model with time coupling, thus allowing aggregators to define their cost coefficients directly based on the feasible regions determined by their flexibility models.
  \item Based on the flexibility cost quantification method, we propose a DSO optimization model for coordinating the flexibility of all aggregators in the distribution system to participate in the transmission system operation for energy arbitrage and ancillary services provision.
  \item From the DSO optimization model, we derive the concept of \emph{marginal flexibility prices} for quantifying the marginal contributions of each aggregator's flexibility to the total revenue of the DSO. The marginal flexibility prices are then used to allocate the revenues earned from the transmission system to each aggregator, ensuring a non-profit DSO in the flexibility activation process. 
\end{itemize}


\emph{Notations:} Bold letters (e.g., $\mathbf x$, $\mathbf A$) denote vectors and matrices, and the corresponding regular letters with subscripts represent their components. The operator $(\cdot)^{\top}$ calculates the vector/matrix transposition, and the operator $|(\cdot)|$ calculates the cardinality of a set.
\vspace{-7pt}
\section{Quantification of Flexibility Costs for Individual DERs}\label{Sect:flexibility_cost}
This section first presents the flexibility cost quantification method for individual DERs based on adjustment ranges in both power and accumulated energy consumption. Then, we discuss how to specify the cost coefficients in this method according to the distinct characteristics of various DERs. 
\vspace{-10pt}
\subsection{Flexibility Cost Quantification based on Both Power and Energy Adjustment Ranges}
When DERs provide flexibilities to the power system, their power may be adjusted within the offered flexibility region. Power adjustments can lead to equipment degradation or inconvenience for users, which constitute the primary source of the DERs' flexibility costs. Flexibility costs are considered in the pre-operational decision-making stage, which reflects the expected adjustment costs in the subsequent real-time operation stage, so they should be quantified based on the adjustment capability defined by the flexibility region. 

Conceptually, flexibility costs are similar to the capacity costs for generators providing ancillary services.\footnote{In ancillary services, mileage costs may also be counted, but in a post-operational fashion. Here, the primary purpose of formulating the flexibility cost is to support pre-operational decision-making. In this context, mileage costs are typically estimated by multiplying capacity costs by a proportion based on historical regulation data \cite{mousavi2021dso}. Therefore, we focus on capacity costs.} However, since many DERs' flexibility region exhibits time-coupled characteristics, their flexibility costs can not be adequately captured based on the power adjustment range of each time slot, which is, though, suitable for conventional generators. Hence, we propose to quantify the individual DERs' flexibility costs based on adjustment ranges in both power and accumulated energy, which aligns with the classic power-energy flexibility model (with time-coupling) of individual DERs \cite{xu2016hierarchical}. 

First, the operational time horizon is discretized into $T$ slots, denoted as $[T]\triangleq\{1,2,..., T\}$. Each time slot has a length of $\Delta_T$ and is indexed by $t$. The flexibility cost of one DER is:
\begin{equation}
  C_{\text{DER}} = {\hat{\mathbf{c}}_{\text{p}}^{\top} \Delta \hat{\mathbf{p}}  + \check{\mathbf{c}}_{\text{p}}^{\top} \Delta \check{\mathbf{p}}   + \hat{\mathbf{c}}_{\text{e}}^{\top} \Delta \hat{\mathbf{e}}   + \check{\mathbf{c}}_{\text{e}}^{\top} \Delta \check{\mathbf{e}} }  ,\label{eq:flex_cost}
\end{equation}
where $\Delta \hat{\mathbf{p}} \in \mathbb{R}^T_+$ and $\Delta \check{\mathbf{p}} \in \mathbb{R}^T_+$ denote the vectors of activated upward and downward adjustment ranges of the power profile $\mathbf{p} \in \mathbb{R}^T$ relative to the baseline profile $\mathbf{p}^{{\text{base}}}\in \mathbb{R}^T$, respectively; $\Delta \hat{\mathbf{e}} \in \mathbb{R}^T_+$ and $\Delta \check{\mathbf{e}} \in \mathbb{R}^T_+$ represent the activated\footnote{Because not all the flexibilities offered by the DER are valuable for the system, only a portion of the flexibility will be actually utilized in the power system operations. This paper calls this portion of the DER's flexibility "activated."} upward and downward adjustment ranges of the accumulated energy consumption trajectory $\mathbf{e}\in \mathbb{R}^T$ relative to the baseline energy trajectory $\mathbf{e}^{{\text{base}}}\in \mathbb{R}^T$, respectively; $\hat{\mathbf{c}}_{\text{p}} \in \mathbb{R}^T$, $\check{\mathbf{c}}_{\text{p}} \in \mathbb{R}^T$, $\hat{\mathbf{c}}_{\text{e}} \in \mathbb{R}^T$, and $\check{\mathbf{c}}_{\text{e}} \in \mathbb{R}^T$ denote the vectors of cost coefficients for upward and downward adjustment ranges in power and energy, depending on the specific DERs. Variables $\mathbf{p}$, $\Delta \hat{\mathbf{p}}$, $\Delta \check{\mathbf{p}}$, $\mathbf{e}$, $\Delta \hat{\mathbf{e}}$, and $\Delta \check{\mathbf{e}}$ should fulfill the following constraints:
\begin{equation}
  \underline{\mathbf{p}} \le \mathbf{p}^{{\text{base}}} - \Delta \check{\mathbf{p}}  \le \mathbf{p} \le \mathbf{p}^{{\text{base}}} + \Delta \hat{\mathbf{p}}\le \overline{\mathbf{p}}  ,\label{eq:power_deviation}
\end{equation}
\begin{equation}
  \underline{\mathbf{e}} \le \mathbf{e}^{{\text{base}}}- \Delta \check{\mathbf{e}}\le \mathbf{e}  \le  \mathbf{e}^{{\text{base}}}+\Delta \hat{\mathbf{e}}\le \overline{\mathbf{e}}, \label{eq:energy_deviation}
\end{equation}
\begin{equation}
  e_t^{{\text{base}}} = \sum\limits_{\tau  = 1}^t {p_\tau ^{{\text{base}}}\Delta_T} ,e_t = \sum\limits_{\tau  = 1}^t {p_\tau \Delta_T},\forall t \in [T],\label{eq:power_energy_relationship}
\end{equation}
where $\overline{\mathbf{p}}\in \mathbb{R}^T$, $\underline{\mathbf{p}}\in \mathbb{R}^T$, $\overline{\mathbf{e}}\in \mathbb{R}^T$, and $\underline{\mathbf{e}}\in \mathbb{R}^T$ denotes the DER's physical upper and lower bounds of power and accumulated energy. Constraints \eqref{eq:power_deviation} and \eqref{eq:energy_deviation} restrict the DER's power and energy profile within their activated adjustment ranges and the activated power and energy adjustment ranges within their physical bounds. Equation \eqref{eq:power_energy_relationship} gives the relationship between power and energy.


The constraints in \eqref{eq:power_deviation}-\eqref{eq:power_energy_relationship} actually constitute the power-energy boundary model that describes the flexibility of typical DERs such as PVs, EVs, BESSs, and HPs, with bound parameters determined by their operational characteristics, as detailed in \cite{xu2016hierarchical}. The formulation \eqref{eq:flex_cost} aligns with the flexibility model \eqref{eq:power_deviation}-\eqref{eq:power_energy_relationship} since it calculates the flexibility cost based on the activated adjustment ranges in both power and accumulated energy consumption. By properly setting the cost coefficient vectors $\hat{\mathbf{c}}_{\text{p}} $, $\check{\mathbf{c}}_{\text{p}}$, $\hat{\mathbf{c}}_{\text{e}}$, and $\check{\mathbf{c}}_{\text{e}}$, formula \eqref{eq:flex_cost} can reflect the costs of DERs offering flexibility. The next subsection will discuss the specification of these coefficients for typical resources.
\vspace{-10pt}
\subsection{Specification of the Cost Coefficients for Typical DERs}
For resources whose flexibility can be described by the power adjustment range in each time slot (such as distributed PV and curtailable loads), flexibility cost can also be quantified using power adjustment ranges in each time slot. Therefore, the specification of the cost coefficients becomes straightforward: set $\hat{\mathbf{c}}_{\text{e}}$ and $\check{\mathbf{c}}_{\text{e}}$ to zero and specify $\hat{\mathbf{c}}_{\text{p}}$ and $\check{\mathbf{c}}_{\text{p}}$ using the same manner as traditional power-range-based formulations.

For resources with batteries such as EVs and BESSs, providing flexibility may make their charging and discharging process repeated or discontinuous, leading to battery aging. Given the energy trajectory, the battery aging can be estimated by the rainflow counting method \cite{Shi2019Optimal,Diao2024Subgradient}. Since an EV or BESS providing flexibility means that its energy trajectory can be adjusted within the energy adjustment range, the device degradation part of its flexibility cost can be estimated by counting the expected battery aging cost on typical control signals, and the values of $\hat{\mathbf{c}}_{\text{e}}$ and $\check{\mathbf{c}}_{\text{e}}$ can be set accordingly.

In addition to the equipment degradation costs, flexibility costs also include the inconvenience caused to the DER users by the operational adjustment. This aspect can be directly perceived by DER users when they offer flexibility to the power system, yet it has not been adequately discussed in the literature. Therefore, the subsequent part of this subsection focuses on how to specify the flexibility cost coefficients for storage-like DERs, including EVs, BESS, and HPs, based on the potential inconvenience caused to their users.

\subsubsection{Electric Vehicles}\;

The flexibility of EVs lies in the charging process during their plug-in periods. Since the primary goal of EV charging is to charge the battery to the expected energy level by departure, the inconvenience caused to EV users by operational adjustment is reflected in the potentially unsatisfied charging demand at the departure time. 

\begin{figure}[!t]
  \centering
  \includegraphics[width=2.8in]{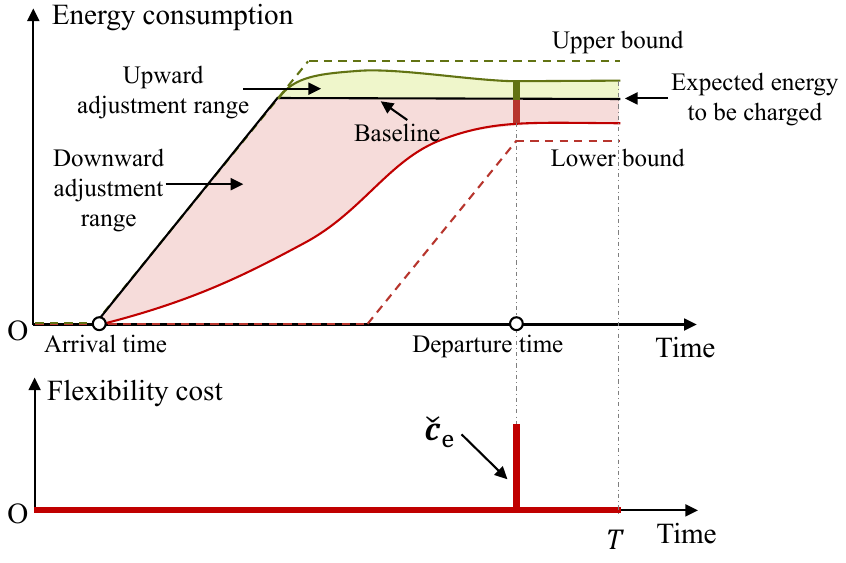}
  \caption{The flexibility in accumulated energy consumption of an EV and its associated cost coefficients.}
  \label{fig:EV}
  \vspace{-15pt}
\end{figure}

Fig. \ref{fig:EV} illustrates a typical EV's adjustment ranges in the accumulated energy consumption and its flexibility cost coefficients. In this example, the EV has an expected value for the energy to be charged by departure but also allows it to be adjustable within certain upper and lower bounds (the green and red dashed lines). The upper bound typically corresponds to the battery's capacity limit, while the lower bound represents the minimum energy level acceptable to the EV user. The detailed calculation process of the upper and lower energy bound trajectories of EVs, as well as the subsequent BESSs and HPs, can be found in \cite{xu2016hierarchical} and is omitted in this paper. The baseline energy trajectory is the energy trajectory where charging starts immediately at the rated power upon plug-in until the expected energy level is reached. The area within the upward and downward energy adjustment ranges represents the activated flexibility in accumulated energy consumption, which is where the actual energy trajectory can be adjusted. However, only the energy adjustment at the departure time can cause inconvenience to the EV user. More precisely, it is the downward adjustment that matters, as the upward adjustment, representing charging more than the expected energy, typically does not lead to an inconvenience in the EV's usage. Thus, in the flexibility cost coefficients, only the coefficient for the downward energy adjustment range $\check{\mathbf{c}}_{\text{e}}$ at the departure time is non-zero, while $\check{\mathbf{c}}_{\text{e}}$ at other times, $\hat{\mathbf{c}}_{\text{e}}$, $\check{\mathbf{c}}_{\text{p}}$ and $\hat{\mathbf{c}}_{\text{p}}$ are all zero.\footnote{The increased energy cost due to the overcharging indicated by the upward adjustments in the energy trajectory is not considered within the scope of flexibility costs. Instead, these costs will be incorporated into the energy cost component in the objective function of the flexibility activation program.}


In summary, based on the flexibility cost model \eqref{eq:flex_cost}, the aggregator can establish a contract with EV users that specifies compensation for any unsatisfied charging demands at the departure time. This contract directly correlates to the value of \(\check{\mathbf{c}}_{\text{e}}\). Consequently, EV users can intuitively understand the potential inconvenience to their EV usage due to operational adjustment and assess their expected benefits. This level of clarity and directness is not achievable with traditional cost quantification methods based on power adjustment ranges, as power adjustments during charging are invisible to EV users.

\subsubsection{Battery Energy Storage Systems}\;

The flexibility of a BESS lies in its ability to charge and discharge within the battery's capacity. The energy level of a BESS typically needs periodic balancing to preserve sufficient capacity for potential charging and discharging demands. Therefore, the inconvenience caused to a BESS user by operational adjustment can be reflected by the unbalanced energy at times of balancing.

\begin{figure}[!t]
  \centering
  \includegraphics[width=2.8in]{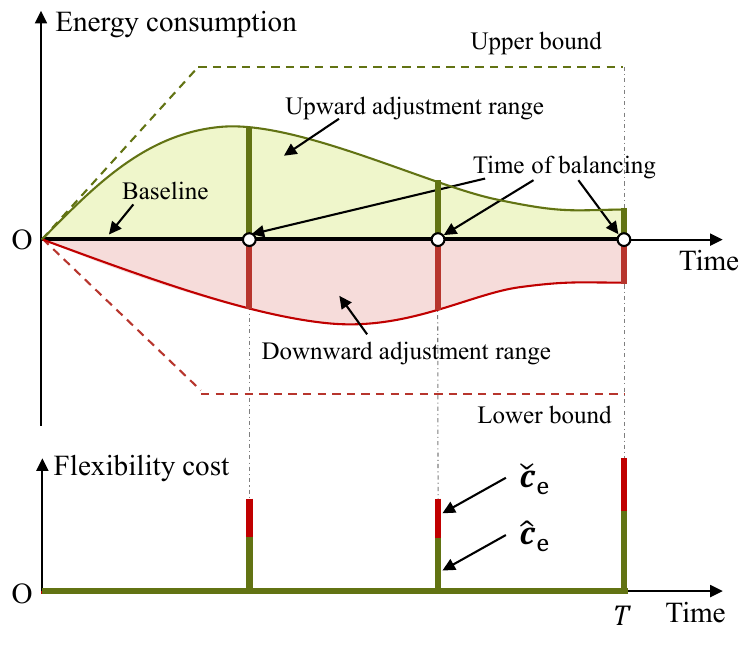}
  \caption{The flexibility in accumulated energy consumption of a BESS and its associated cost coefficients.}
  \label{fig:BESS}
  \vspace{-15pt}
\end{figure}

Fig. \ref{fig:BESS} illustrates a typical BESS's energy adjustment range and the corresponding cost coefficients. Three balancing points are set in the decision time horizon, at which the BESS user receives compensation if its energy does not return to the initial level. This compensation is translated into the settings of cost coefficients, where both $\hat{\mathbf{c}}_{\text{e}}$ and $\check{\mathbf{c}}_{\text{e}}$ are nonzero at the pre-defined balancing times. The cost coefficient for the downward energy adjustment range $\check{\mathbf{c}}_{\text{e}}$ is set larger than that for upward capacity $\hat{\mathbf{c}}_{\text{e}}$, indicating that compensation for insufficient energy is larger than that for overcharged energy. Cost coefficients at the end of the time horizon $T$ are higher than those at other balancing times, indicating the highest priority of the energy balancing across the whole time horizon. This balance may also be imposed as a hard constraint by setting the upper and lower energy bounds at the end of the time horizon to zero, ensuring the battery's energy finally returns to the initial level.

\subsubsection{Heat Pumps}\;

HPs can offer flexibility to the power system due to the allowable adjustment range in the buildings' indoor temperature. Hence, temperature adjustment ranges associated with the activated flexibility most appropriately measure the potential inconvenience caused to HP users. The indoor temperature evolution of a building with an HP can be approximately described using the linear thermal dynamic model \cite{you2024leveraging}:
\begin{equation}
  {\theta _t} = \alpha {\theta _{t - 1}} + (1 - \alpha )(\theta _t^{{\text{amb}}} + \frac{{\eta }}{H} \cdot {p_t}),\forall t \in [T],\label{eq:temp_evolution}
\end{equation}
where ${\theta _t}$ denotes the indoor temperature in Kelvin (K) in the $t$-th time slot, $\alpha = \exp(-{\Delta_T H}/{C})$, $C$ and $H$  are the thermal capacitance (kWh/K) and resistance (K/kW), $\theta _t^{{\text{amb}}}$ represents the ambient temperature, and $\eta$ is the coefficient of performance (COP) of the HP (assumed constant).

We define the baseline power and accumulated energy consumption profiles, $\mathbf p ^{\text{base}}$ and $\mathbf e ^{\text{base}}$
, as those that always maintain the indoor temperature at the set point \(\theta^\text{set}_t, \forall t\in [T]\). Activating the flexibility of HPs leads to the fluctuation of indoor temperature within the upward and downward adjustment ranges relative to the temperature set point, denoted by $\Delta \hat{\theta}_t$ and $\Delta \check{\theta}_t$, respectively, (with vector forms $\Delta \hat{\boldsymbol{\theta}} \in \mathbb{R}^T_+$ and $\Delta \check{\boldsymbol{\theta}} \in \mathbb{R}^T_+$). Regarding the relationship between the temperature adjustment ranges, $\Delta \hat{\boldsymbol{\theta}}$ and $\Delta \check{\boldsymbol{\theta}}$, and the upward and downward adjustment ranges in power, $\Delta \hat{\mathbf{p}}$ and $\Delta \check{\mathbf{p}}$, and accumulated energy consumption, $\Delta \hat{\mathbf{e}}$ and $\Delta \check{\mathbf{e}}$, the following Proposition holds:
\newtheorem{proposition}{Proposition}
\begin{proposition}
  $ \exists \mathbf D \in \mathbb R^{T\times T}$, $\forall \Delta \hat{\boldsymbol{\theta}}\ge \mathbf 0 $ and $\Delta \check{\boldsymbol{\theta}}\ge \mathbf 0$, if the indoor temperature profile $\boldsymbol{\theta}$ satisfies $-\Delta \check{\boldsymbol{\theta}} \le \Delta \boldsymbol{\theta}-\boldsymbol{\theta}^{\text{set}}\le \hat{\boldsymbol{\theta}}$, then the corresponding accumulated energy consumption profile $\mathbf e$ satisfies $-\Delta \check{\mathbf{e}}\le \mathbf{e} - \mathbf e ^{\text{base}} \le \Delta \hat{\mathbf{e}}$, where
  \begin{equation}
    \Delta \hat{\mathbf{e}} = {\mathbf{D}}\Delta \hat{\boldsymbol{\theta}}, \;\;\Delta \check{\mathbf{e}} = {\mathbf{D}}\Delta \check{\boldsymbol{\theta}}. \label{eq:HP_T2E}
  \end{equation}
Specifically, $\mathbf D $ is an invertible matrix defined as:
\begin{equation}
  {\mathbf{D}} \triangleq \frac{\Delta_T H}{\eta(1 - \alpha ) }\left[ {\begin{array}{cccc}
    1&{}&{}&{}\\
    {1 - \alpha }&1&{}&{}\\
     \vdots & \ddots & \ddots &{}\\
    {1 - \alpha }& \cdots &{1 - \alpha }&1
    \end{array}} \right].\nonumber
\end{equation} 
However, $ \nexists \mathbf F \in \mathbb R^{T\times T}$, such that $\forall \Delta \hat{\boldsymbol{\theta}}\ge \mathbf 0 $ and $\Delta \check{\boldsymbol{\theta}}\ge \mathbf 0$, if $-\Delta \check{\boldsymbol{\theta}} \le \boldsymbol{\theta}-\boldsymbol{\theta}^{\text{set}}\le \Delta \hat{\boldsymbol{\theta}}$, then the corresponding power profile $\mathbf p$ satisfies  $-\mathbf F \Delta \check{\boldsymbol{\theta}}\le \mathbf{p} - \mathbf p ^{\text{base}} \le \mathbf F \Delta \hat{\boldsymbol{\theta}}$.
\end{proposition}

\emph{Proof:} See Appendix.

Proposition 1 indicates that the upward and downward temperature adjustment ranges $\Delta \hat{\boldsymbol{\theta}}$ and $\Delta \check{\boldsymbol{\theta}}$ can be linearly transformed into the upward and downward adjustment ranges in accumulated energy consumption $\Delta \hat{\mathbf{e}}$ and $\Delta \check{\mathbf{e}}$, respectively. However, $\Delta \hat{\boldsymbol{\theta}}$ and $\Delta \check{\boldsymbol{\theta}}$ cannot be respectively transformed into upward and downward power adjustment ranges $\Delta \hat{\mathbf{p}}$ and $\Delta \check{\mathbf{p}}$. The rationale behind Proposition 1 can be intuitively understood: the indoor temperature is primarily influenced by the accumulated power over time rather than the power in each time slot itself. 

The transformation in \eqref{eq:HP_T2E} allows aggregators to have a contract with HP users specifying the compensation for per unit temperature adjustment range and to transform this compensation into cost coefficients of energy adjustment range. We denote by \(\hat{\boldsymbol{\rho}} \in \mathbb{R}^T_+\) and \(\check{\boldsymbol{\rho}} \in \mathbb{R}^T_+\) the vectors of compensations for the activated upward and downward temperature adjustment ranges, respectively. Then, the total cost of the activated flexibility of an HP is:
\begin{equation}
  C_{\text{HP}} = \hat{\boldsymbol{\rho}}^{\top}\Delta \hat{\boldsymbol{\theta}}  + \check{\boldsymbol{\rho}}^{\top}\Delta \check{\boldsymbol{\theta}},
\end{equation}
which can then be equivalently reformulated based on energy adjustment ranges:
\begin{equation}
  C_{\text{HP}} = \hat{\boldsymbol{\rho}}^{\top} \mathbf{D}^{-1} \Delta \hat{\mathbf{e}}  + \check{\boldsymbol{\rho}}^{\top} \mathbf{D}^{-1} \Delta \check{\mathbf{e}},
\end{equation}
aligning with the flexibility cost model \eqref{eq:flex_cost}, where $\hat{\mathbf{c}}_{\text{p}} = \check{\mathbf{c}}_{\text{p}}= \mathbf{0}$, $\hat{\mathbf{c}}_{\text{e}}=  (\mathbf{D}^{-1})^{\top}\hat{\boldsymbol{\rho}}$, and $\check{\mathbf{c}}_{\text{e}} = (\mathbf{D}^{-1})^{\top}\check{\boldsymbol{\rho}}$.

\begin{figure}[!t]
  \centering
  \includegraphics[width=2.8in]{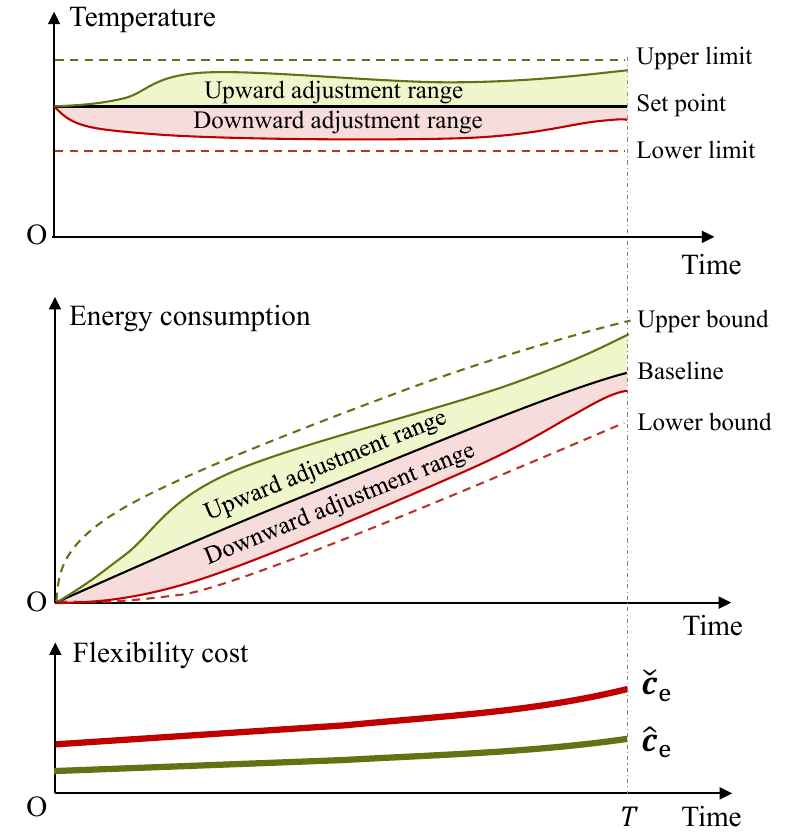}
  \caption{The flexibility in accumulated energy consumption of an HP and its associated cost coefficients.}
  \label{fig:HP}
  \vspace{-15pt}
\end{figure}

Fig. \ref{fig:HP} illustrates adjustment ranges in the indoor temperature, the corresponding adjustment ranges in accumulated energy consumption, and cost coefficients. Cost coefficients of energy adjustment ranges shown in this figure, which increase with time, are generated based on a constant cost coefficient of temperature adjustment range across the time horizon. Take $\check{\mathbf{c}}_{\text{e}}$ as an example, assuming that $\check{\rho}_t = \check{\rho}_0,\forall t\in [T]$, then $\check{\mathbf{c}}_{\text{e}} = ([\check{\rho}_0, \check{\rho}_0, \cdots, \check{\rho}_0]\mathbf{D}^{-1})^{\top} = \frac{\eta(1-\alpha)}{H\Delta_T}\check{\rho}_0[{\alpha ^{T-1}},...,{\alpha ^2},\alpha ,1]^{\top}$. As $0<\alpha<1$ by definition, $\check{c}_{\text{e},t}$ increases with the time index $t$. Besides, $\check{\mathbf{c}}_{\text{e}}$ is typically larger than $\hat{\mathbf{c}}_{\text{e}}$ because downward temperature adjustments cause more discomfort to the people inside the building than upward adjustments considering the primary task of the HP is heating.

Finally, we note that \(\hat{\boldsymbol{\rho}}\) and \(\check{\boldsymbol{\rho}}\) should not be uniform across all buildings, as the thermal dynamic parameters and HPs' COPs vary. Identical temperature adjustments can lead to different energy adjustments in different buildings, so applying uniform \(\hat{\boldsymbol{\rho}}\) and \(\check{\boldsymbol{\rho}}\) would be unfair to HP users in a well-insulated building. Thus, \(\hat{\boldsymbol{\rho}}\) and \(\check{\boldsymbol{\rho}}\) should be customized based on the thermal dynamic parameters of the building and the HP's COP. We suggest setting \(\hat{\boldsymbol{\rho}}\) and \(\check{\boldsymbol{\rho}}\) proportional to \(\frac{H\Delta_T}{\eta(1-\alpha)}\), which aligns the HP's flexibility cost coefficients with the contribution to energy adjustment ranges.

\section{Model and Costs of Aggregated Flexibility}\label{sect:agg_flex_price}
We have discussed the flexibility cost formulation of individual DERs. This section focuses on the aggregator's perspective, namely the model and cost for the aggregated flexibility, which the aggregators should report to the DSO. We denote the set of DERs in the aggregator as \(\mathcal K\), indexed by \(k\), and the power vector of the \(k\)-th DER as \(\mathbf{p}_k \in \mathbb{R}^T\). Hence, the aggregated power is \(\mathbf{P} \triangleq \sum\nolimits_{k\in\mathcal K}\mathbf{p}_k\). The aggregated flexibility model refers to the feasible region of \(\mathbf{P}\). Given the individual flexibility constraints \eqref{eq:power_deviation}-\eqref{eq:power_energy_relationship}, the aggregated flexibility, as derived in our previous works \cite{wen2022aggregate} and \cite{wen2023improved}, can be formulated as follows:
\begin{equation}
  \underline {\boldsymbol{\phi }}  \le {\mathbf{UP}} \le \overline {\boldsymbol{\phi }},\label{eq:agg_flex}
\end{equation}
where $\mathbf U$ is an ${M\times T}$ coefficient matrix with each row vector being a ${1\times T}$ vector composed of zeros and ones, and \(M\) represents the number of constraints that is dependent on whether an exact or an approximate model is used. Constraints \eqref{eq:agg_flex} define a polytope in the $\mathbb{R}^T$ space with a specific shape. We let \(\mathbb{B}\triangleq\{0,1\}\) and $\mathbb{B}^T$ denote the set of all $T$-dimensional row vectors composed solely of 0s and 1s. If the row vectors of \(\mathbf U\) span \(\mathbb{B}^T\) except for $\mathbf 0$, i.e., \(M = 2^T - 1\), \(\mathbf U\) corresponds to the exact aggregation model. When the row vector set of \(\mathbf U\) is a subset of \(\mathbb{B}^T\), \(\mathbf U\) corresponds to an approximate model. Bound parameters \(\underline {\boldsymbol{\phi }} \in\mathbb{R}^{M}\) and \(\overline {\boldsymbol{\phi }} \in\mathbb{R}^{M}\) are derived from all the individual power and energy bound parameters in \eqref{eq:power_deviation}-\eqref{eq:energy_deviation}. Reducing the number of rows of \(\mathbf U\) creates an approximate model with low complexity, and appropriately adjusting \(\underline {\boldsymbol{\phi }}\) and \(\overline {\boldsymbol{\phi }}\) forms either an outer or inner approximation. For example, setting the coefficient matrix \(\mathbf U\) to only include power and accumulated energy constraints and the bound parameters \(\underline {\boldsymbol{\phi }}\) and \(\overline {\boldsymbol{\phi }}\) to the sum of all individual power and energy bounds yields an outer approximation \cite{wen2022aggregate}, and properly shrinking the gap between \(\underline {\boldsymbol{\phi }}\) and \(\overline {\boldsymbol{\phi }}\) forms an inner approximation \cite{wen2023improved}. Since the primary focus of this paper is on the quantification of flexibility costs and values rather than on flexibility aggregation modeling, we directly employ the inner approximated aggregation model proposed in our previous research \cite{wen2023improved}, which has been validated for its advanced accuracy and computational efficiency.

We are now ready to introduce the flexibility cost formulation for the aggregated flexibility model \eqref{eq:agg_flex}. For the sake of clarity, we hereafter refer to the product of \(\mathbf{U}\) and the aggregated power profile \(\mathbf{P}\), i.e., \(\mathbf{UP}\), as the \emph{flexibility trajectory} and the product of \(\mathbf{U}\) and the aggregated baseline power profile \(\mathbf{P}^{\text{base}}\), i.e., \(\mathbf{UP}^{\text{base}}\), as the \emph{baseline flexibility trajectory}. Similar to the flexibility model \eqref{eq:power_deviation}-\eqref{eq:power_energy_relationship} for individual DERs, activating the aggregated flexibility means that the flexibility trajectory \(\mathbf{UP}\) can deviate within a certain range relative to the baseline flexibility trajectory \(\mathbf{UP}^{\text{base}}\), which can be formulated as:
\begin{equation}
  \underline {\boldsymbol{\phi }} \le \mathbf{UP}^{\text{base}} - \Delta \check{\boldsymbol{\phi }}  \le \mathbf{UP} \le \mathbf{UP}^{{\text{base}}} + \Delta \hat{\boldsymbol{\phi }}\le \overline{\boldsymbol{\phi }}  ,\label{eq:flexibility_trajectory_deviation}
\end{equation}
where \(\Delta \hat{\boldsymbol{\phi }}\in \mathbb{R}^M_+\) and \(\Delta \check{\boldsymbol{\phi }}\in \mathbb{R}^M_+\) are the activated upward and downward adjustment ranges, respectively. The utility function of the aggregator is defined as a linear function with the activated upward and downward adjustment ranges, that is
\begin{equation}
  C_{\text{agg}} = \hat{\mathbf{c}}_{\text{agg}}^{\top}\Delta \hat{\boldsymbol{\phi }} + \check{\mathbf{c}}_{\text{agg}}^{\top}\Delta \check{\boldsymbol{\phi }},
\end{equation}
where \(\hat{\mathbf{c}}_{\text{agg}}\) and \(\check{\mathbf{c}}_{\text{agg}}\) in \(\mathbb{R}^M_+\) denote the upward and downward cost coefficients for flexibility adjustment ranges, respectively, which the aggregator reports to the DSO.

The next question is how the aggregator can calculate the cost coefficients $\hat{\mathbf{c}}_{\text{agg}}$ and $\check{\mathbf{c}}_{\text{agg}}$. Principally, $\hat{\mathbf{c}}_{\text{agg}}$ and $\check{\mathbf{c}}_{\text{agg}}$ are related to the aggregator's cost of purchasing flexibility from the DER users and its operation costs. Aggregators may develop various strategies to determine $\hat{\mathbf{c}}_{\text{agg}}$ and $\check{\mathbf{c}}_{\text{agg}}$ for reporting to the DSO. Since the focus of this paper is primarily on the quantification of flexibility costs, we here use a simple approach to estimate $\hat{\mathbf{c}}_{\text{agg}}$ and $\check{\mathbf{c}}_{\text{agg}}$ without a detailed discussion of the aggregators' strategies.

First, it should be noted that the flexibility model \eqref{eq:agg_flex} also includes constraints of power and accumulated energy consumption of the aggregator. In other words, the power and accumulated energy consumption are a part of the flexibility trajectory \(\mathbf{UP}\). The power and accumulated energy part of cost coefficients $\hat{\mathbf{c}}_{\text{agg}}$ and $\check{\mathbf{c}}_{\text{agg}}$ are estimated via a weighted average method, with the weights being the ratio of the maximum power and energy adjustment ranges of each DER, i.e.,
\begin{align}
  \hat{c}_{\text{agg},m}  = \frac{{\sum\nolimits_{k \in {{\mathcal K}}} {\hat{c}_{k,m} \left( {{{\overline \phi  }_{k,m}} - \phi _{k,m}^{{\text{base}}}} \right)} }}{{\sum\nolimits_{k \in {{\mathcal K}}} {\left( {{{\overline \phi  }_{k,m}} - \phi _{k,m}^{{\text{base}}}} \right)} }},\forall m\in \mathcal{I}_{\text{p,e}},\\
  \check{c}_{\text{agg},m}  = \frac{{\sum\nolimits_{k \in {{\mathcal K}}} {\check{c}_{k,m} \left( {\phi _{k,m}^{{\text{base}}} - {{\underline \phi  }_{k,m}}} \right)} }}{{\sum\nolimits_{k \in {{\mathcal K}}} {\left( {\phi _{k,m}^{{\text{base}}} - {{\underline \phi  }_{k,m}}} \right)} }},\forall m\in \mathcal{I}_{\text{p,e}}
\end{align}
where \(k\) is the index of DERs and \(\mathcal K\) is the set of all DERs within the aggregator; \(m\) is the index of constraints in the aggregated flexibility model, and \(\mathcal{I}_{\text{p,e}}\) is the indices set of constraints on power and accumulated energy consumption, $|\mathcal{I}_{\text{p,e}}|\le M$. The parameter \(\phi _{k,m}^{{\text{base}}}\) denotes the baseline of power or accumulated energy consumption of the \(k\)-th DER, depending on whether \(m\) corresponds to a constraint on power or accumulated energy consumption. Similarly, \(\overline{\phi} _{k,m}\) and \(\underline{\phi} _{k,m}\) respectively denote the upper and lower bounds of power or accumulated energy consumption, and \(\hat{c}_{k,m}\) and \(\check{c}_{k,m}\) are the cost coefficients of the adjustment ranges in power or accumulated energy consumption specified in Section \ref{Sect:flexibility_cost}.

If the aggregated flexibility model \eqref{eq:agg_flex} includes constraints beyond power and accumulated energy consumption, i.e., $M>|\mathcal{I}_{\text{p,e}}|$, the cost coefficients of adjustment ranges in those additional parts should be directly set to zero. This setting avoids the duplication of the cost calculation, as power and accumulated energy adjustments could lead to adjustments in other segments of the flexibility trajectory. This does not imply that other segments of the flexibility trajectory do not contribute to the flexibility cost of the aggregator since those segments are subject to constraints and indirectly influence the aggregator's total flexibility cost \(C_{\text{agg}}\). Indeed, for simplicity, the DSO can require aggregators to report a flexibility model with only power and accumulated energy constraints and to report the cost coefficients only for adjustment ranges in this segment. Nevertheless, since we have designed a more general flexibility cost formulation, it holds the potential to be applied to more accurate flexibility models.

\section{The DSO Model for Flexibility Activation and Revenue Allocation}
The DSO optimizes the activation of flexibilities from aggregators for energy arbitrage and ancillary service provision at the transmission level. Here, we primarily consider reserve capacities provision as the ancillary service for participation, while the proposed model also adapts to other ancillary services. In this context, the DSO should 1) determine the reference power profile and reserve capacities offered to the TSO, 2) determine the activated adjustment range for the flexibility trajectory of each aggregator, and 3) allocating the revenues obtained from the TSO to each aggregator.

\vspace{-5pt}
\subsection{The DSO Optimization Model for Flexibility Activation}
The objective of the DSO is to minimize the net cost of the distribution system, formulated as:
\begin{align}
  \min C_{\text{net}} &= {C_{{\text{Energy}}}} - {R_{{\text{Capacity}}}} + {C_{{\text{Flexibility}}}},\label{eq:market_obj}\\
    {C_{{\text{Energy}}}} &\triangleq {{\mathbf{c}}_{\text{energy}}^{\top}}{{\mathbf{P}}^{\text{ref}}_0},\nonumber\\
    {R_{\text{Capacity}}} &\triangleq  {\mathbf{c}}_{\text{ru}}^{\top}{{\mathbf{R}}^{\text{up}}} + {\mathbf{c}}_{\text{rd}}^{\top}{{\mathbf{R}}^{\text{dn}}},\nonumber\\
    {C_{{\text{Flexibility}}}} &\triangleq  \sum\nolimits_{h \in \mathcal A} {(\hat{\mathbf{c}}_h^{\top}\Delta \hat{\boldsymbol{\phi }}_h + \check{\mathbf{c}}_h^{\top}\Delta \check{\boldsymbol{\phi }}_h)} ,\nonumber
\end{align}
where ${C_{{\text{Energy}}}}$ is the total energy cost of the distribution system, ${R_{\text{Capacity}}}$ is the revenue from providing reserve capacities, and ${C_{{\text{Flexibility}}}}$ is the total flexbility cost; ${\mathbf{c}}_{\text{energy}} \in \mathbb{R}^T$, ${\mathbf{c}}_{\text{ru}} \in \mathbb{R}^T$, and ${\mathbf{c}}_{\text{rd}} \in \mathbb{R}^T$ denote the vectors composed of the energy price, up-reserve price and down-reserve price, respectively, at the transmission level over the time horizon $[T]$ (assuming that the DSO acts as a price taker at the transmission system, ${\mathbf{c}}_{\text{energy}}$ , ${\mathbf{c}}_{\text{ru}}$, and ${\mathbf{c}}_{\text{rd}} $ are input parameters at the DSO level); ${{\mathbf{P}}^{\text{ref}}_0} \in \mathbb{R}^T$ represents the reference power profile at the TSO-DSO interface, i.e., the root node of the distribution system, while \({{\mathbf{R}}^{\text{up}}} \in \mathbb{R}^T_+\) and \({{\mathbf{R}}^{\text{dn}}} \in \mathbb{R}^T_+\) represent the up- and down-reserve capacities offered to the TSO, respectively; aggregators in the distribution system are indexed by \(h\), with the set of all aggregators denoted as \(\mathcal A\); \(\Delta \hat{\boldsymbol{\phi }}_h\) and \(\Delta \check{\boldsymbol{\phi }}_h\) denote the activated upward and downward adjustment range of the flexibility trajectory of aggregator \(h\), respectively, and \(\hat{\mathbf{c}}_h\) and \(\check{\mathbf{c}}_h\) denote the corresponding flexibility cost coefficients, as defined in Section \ref{sect:agg_flex_price}.

The DSO's reserve capacity offered to the TSO must ensure that any adjustment signal within this capacity is feasible with respect to both the distribution network constraints and aggregators' flexibility constraints, which means the following condition should be satisfied:
\begin{equation}
  {{\mathbf{P}}_0} \in \mathbb{X},\forall {{\mathbf{P}}_0}:{\mathbf{P}}_0^{{\text{ref}}} - {{\mathbf{R}}^{{\text{up}}}} \le {{\mathbf{P}}_0} \le {\mathbf{P}}_0^{{\text{ref}}}+{{\mathbf{R}}^{{\text{dn}}}},\label{eq:condition_feasibility}
\end{equation}
where \(\mathbb{X}\) represents the feasible region of the power profile ${{\mathbf{P}}_0}\in \mathbb{R}^T$ at the root node of the distribution system, determined by the distribution power flow constraints and the flexibility models of all aggregators. The upper and lower bounds of ${{\mathbf{P}}_0}$ are \({\mathbf{P}}_0^{{\text{ref}}}+{{\mathbf{R}}^{{\text{dn}}}}\) and \({\mathbf{P}}_0^{{\text{ref}}} - {{\mathbf{R}}^{{\text{up}}}}\), respectively. Down-reserve is related to the plus sign, and up-reserve is related to the minus sign because reserves are defined from the perspective of generators, whereas \({\mathbf{P}}_0^{{\text{ref}}}\) and ${{\mathbf{P}}_0}$ represent loads with an opposite direction. As we employ the LinDistFlow model to describe the distribution power flow constraints, the closed form of \(\mathbb{X}\) can also be expressed as \eqref{eq:agg_flex}, but the parameters $\overline {\boldsymbol{\phi }}$ and $\underline {\boldsymbol{\phi }}$ will be affected by the distribution network constraints (See our previous work \cite{9851563} for a detailed theoretical analysis). Since all the components in $\mathbf U$ are no less than zero, \({\mathbf{P}}_0^{{\text{ref}}} - {{\mathbf{R}}^{{\text{up}}}} \in \mathbb{X}\) and \({\mathbf{P}}_0^{{\text{ref}}} + {{\mathbf{R}}^{{\text{dn}}}} \in \mathbb{X}\) is sufficient to ensure that any ${{\mathbf{P}}_0}$ such that ${\mathbf{P}}_0^{{\text{ref}}} - {{\mathbf{R}}^{{\text{up}}}} \le {{\mathbf{P}}_0} \le {\mathbf{P}}_0^{{\text{ref}}}+{{\mathbf{R}}^{{\text{dn}}}}$ also belongs to \(\mathbb{X}\). Consequently, the condition \eqref{eq:condition_feasibility} becomes equivalent to \({\mathbf{P}}_0^{{\text{ref}}} - {{\mathbf{R}}^{{\text{up}}}} \in \mathbb{X}\) and \({\mathbf{P}}_0^{{\text{ref}}} + {{\mathbf{R}}^{{\text{dn}}}} \in \mathbb{X}\). For the sake of brevity, we let 
\begin{equation}
  {{\mathbf{P}}^{{\text{ru}}}_0} \triangleq {{\mathbf{P}}^{{\text{ref}}}_0} - {{\mathbf{R}}^{{\text{up}}}}\;\;\;\;\text{and}\;\;\;\;{{\mathbf{P}}^{{\text{rd}}}_0} \triangleq {{\mathbf{P}}^{{\text{ref}}}_0} + {{\mathbf{R}}^{{\text{dn}}}}\label{eq:def_pru_prd}
\end{equation}
denote the power profiles at the root node of the distribution system corresponding to the upper and lower bounds of the reserve capacity, respectively, and let \(\mathcal W\triangleq \{\text{ru,rd}\}\) denote the set of superscripts representing up- and down-reserves. 

Constraints in the distribution system include:
  \begin{align}
    & {\mathbf{P}}_0^\omega  = \sum\nolimits_{j \in 0} {{\mathbf{P}}_{0j}^\omega } ,\forall \omega  \in {{\mathcal W}},\label{eq:LinDistFlow:1}\\
    &\sum\nolimits_{j \in i} {{\mathbf{P}}_{ij}^\omega }  + {\mathbf{P}}_i^{{\text{fix}}} + {\mathbf{P}}_i^{{\text{flex,}}\omega } = {\mathbf{0}},\forall i \in {\mathcal N}_+ ,\omega  \in {{\mathcal W}},\label{eq:LinDistFlow:2}\\
    &\sum\nolimits_{j \in i} {{\mathbf{Q}}_{ij}^\omega }  + {\mathbf{Q}}_i^{{\text{fix}}} + {\mathbf{Q}}_i^{{\text{flex,}}\omega } = {\mathbf{0}},\forall i \in {\mathcal N}_+ ,\omega  \in {{\mathcal W}},\label{eq:LinDistFlow:3}\\
    &{\mathbf{V}}_i^\omega  - {{\mathbf{V}}_{j}^\omega}  = 2({r_{ij}}{{\mathbf{P}}_{ij}^\omega}  + {x_{ij}}{{\mathbf{Q}}_{ij}^\omega} ),\forall ij \in {{\mathcal L}},\omega  \in {{\mathcal W}},\label{eq:LinDistFlow:4}\\
    &{\mathbf{V}}_0^\omega  = {\mathbf{1}},\forall \omega  \in {{\mathcal W}},\label{eq:LinDistFlow:5}\\
    &{\underline {\mathbf{V}} _i} \le {\mathbf{V}}_i^\omega  \le {\overline {\mathbf{V}} _i},\forall i \in {\mathcal N}_+ ,\omega  \in {{\mathcal W}},\label{eq:LinDistFlow:6}\\
    &{\mathbf{P}}_i^{{\text{flex,}}\omega } = \sum\nolimits_{h \in {{{\mathcal A}}_i}} {{\mathbf{P}}_h^\omega } ,\forall i \in {\mathcal N}_+ ,\omega  \in {{\mathcal W}},\label{eq:LinDistFlow:7}\\
    &{\mathbf{Q}}_i^{{\text{flex,}}\omega } = \sum\nolimits_{h \in {{{\mathcal A}}_i}} {{\mathbf{Q}}_h^\omega } ,\forall i \in {\mathcal N}_+ ,\omega  \in {{\mathcal W}},\label{eq:LinDistFlow:8}\\
    &{\mathbf{Q}}_h^\omega  = {\mathbf{P}}_h^\omega \tan {\gamma _h},\forall h \in {{\mathcal A}},\omega  \in {{\mathcal W}},\label{eq:LinDistFlow:9}\\
    &{{\mathbf{U}}_h}{\mathbf{P}}_h^\omega  \le {{\mathbf{U}}_h}{\mathbf{P}}_h^{{\text{base}}} + \Delta \hat{\boldsymbol{\phi }}_h :\hat{\boldsymbol{\lambda}}_h^\omega ,\forall h \in {{\mathcal A}},\omega  \in {{\mathcal W}},\label{eq:LinDistFlow:10}\\    
    &{{\mathbf{U}}_h}{\mathbf{P}}_h^\omega  \ge {{\mathbf{U}}_h}{\mathbf{P}}_h^{{\text{base}}} - \Delta \check{\boldsymbol{\phi }}_h:\check{\boldsymbol{\lambda}}_h^\omega ,\forall h \in {{\mathcal A}},\omega  \in {{\mathcal W}},\label{eq:LinDistFlow:11}    \\
    &{{\mathbf{U}}_h}{\mathbf{P}}_h^{{\text{base}}} + \Delta \hat{\boldsymbol{\phi }}_h  \le {\overline {\boldsymbol{\phi }} _h},\forall h \in {{\mathcal A}},\label{eq:LinDistFlow:12}    \\
    &{{\mathbf{U}}_h}{\mathbf{P}}_h^{{\text{base}}} - \Delta \check{\boldsymbol{\phi }}_h  \ge {\underline {\boldsymbol{\phi }} _h},\forall h \in {{\mathcal A}},\label{eq:LinDistFlow:13}   \\
    & \Delta \hat{\boldsymbol{\phi }}_h \ge \mathbf{0}, \Delta \check{\boldsymbol{\phi }}_h \ge \mathbf{0},\forall h \in {{\mathcal A}},\label{eq:LinDistFlow:14}
    \end{align}
where $i/\mathcal N$ denotes the index/set of nodes in the distribution system, $0$ represents the root node, and $\mathcal N_+\triangleq \mathcal N \backslash \{0\}$; $\mathcal L$ denotes the set of lines where $ij$ refers to the line between node $i$ and node $j$; $j\in i$ indicates that node $j$ is connected to node $i$; $\mathcal A_i$ denotes the set of aggregators located at node $i$, $\bigcup_{i\in \mathcal N}\mathcal A_i = \mathcal A$; all bold letters are $T\times 1$ vectors except for $\mathbf{U}_h$, \(\hat{\boldsymbol{\phi }}_h\), \(\check{\boldsymbol{\phi }}_h\), ${\overline {\boldsymbol{\phi }} _h}$, and ${\underline {\boldsymbol{\phi }} _h}$; variables with superscript \(\omega\) vary across scenarios of up- and down-reserves; $\mathbf{P}_{ij}^{\omega}/\mathbf{Q}_{ij}^{\omega}$, $\mathbf{P}^{\text{fix}}_{i}/\mathbf{P}^{\text{fix}}_{i}$ and $\mathbf{P}^{\text{flex},\omega}_{i}/\mathbf{Q}^{\text{flex},\omega}_{i}$ denote the active/reactive power flow in line $ij$, fixed load at node $i$, and flexible load at node $i$, respectively; $\mathbf{V}^{\omega}_{i}$ is the square of voltage at node $i$; $r_{ij}$ and $x_{ij}$ denote the resistance and reactance of branch $ij$, respectively; $\overline{\mathbf{V}}_{i}$ and $\underline{\mathbf{V}}_{i}$ denote the square of upper and lower voltage limit of node $i$, respectively; ${{\mathbf{P}}_h^\omega }/{{\mathbf{Q}}_h^\omega }$ is the active/reactive power of aggregator $h$ and $\gamma_h$ denotes the power factor angle of aggregator $h$; and ${{\mathbf{U}}_h}$, ${\mathbf{P}}_h^{{\text{base}}}$, ${\overline {\boldsymbol{\phi }} _h}$, and ${\underline {\boldsymbol{\phi }} _h}$ have been defined in Section \ref{sect:agg_flex_price}, with the aggregator's index $h$ appended here. 
 
Equation \eqref{eq:LinDistFlow:1} calculates the power at the root node. 
Constraints \eqref{eq:LinDistFlow:2}-\eqref{eq:LinDistFlow:4} are the LinDistFlow equations. 
Constraint \eqref{eq:LinDistFlow:5} sets the voltage of the root node to 1 p.u. and \eqref{eq:LinDistFlow:6} restricts the voltage of other nodes in the distribution system. 
Equations \eqref{eq:LinDistFlow:7} and \eqref{eq:LinDistFlow:8} calculate the flexible active and reactive power of node $i$ as the sum of the power of all the aggregators located at this node. 
Constraint \eqref{eq:LinDistFlow:9} gives the relationship between the active and reactive power of aggregators by assuming a constant power factor. 
Constraints \eqref{eq:LinDistFlow:10}-\eqref{eq:LinDistFlow:14} are the aggregated flexibility model, where the flexibility trajectories of both up- and down-reserves, ${{\mathbf{U}}_h}{\mathbf{P}}_h^{\text{ru}}$ and ${{\mathbf{U}}_h}{\mathbf{P}}_h^{\text{rd}}$, are enveloped by the activated adjustment ranges \(\Delta \hat{\boldsymbol{\phi }}_h\) and \(\Delta \check{\boldsymbol{\phi }}_h\).

In summary, the DSO optimization model for flexibility activation is formulated as a linear programming: \eqref{eq:market_obj}, \eqref{eq:def_pru_prd}-\eqref{eq:LinDistFlow:14}. Solving this model yields the reference power profile ${{\mathbf{P}}^{\text{ref}}_0}$ and reserve capacities ${{\mathbf{R}}^{\text{up}}}$ and ${{\mathbf{R}}^{\text{dn}}}$ that the DSO offers to the TSO, as well as the activated flexibility \(\Delta \hat{\boldsymbol{\phi }}_h\) and \(\Delta \check{\boldsymbol{\phi }}_h\) of each aggregator within the distribution system.
\vspace{-5pt}
\subsection{Marginal Flexibility Prices for Revenue Allocation}
Next, the revenues earned from the TSO should be allocated to each aggregator in the distribution system. A reasonable method is to allocate the revenues based on the marginal contributions of each aggregator's activated flexibility to the total revenue of the distribution system, which encourages each aggregator to report its true costs (\emph{incentive compatibility}). We propose the concept of \emph{Marginal Flexibility Prices} (MFPs) to quantify the marginal contributions of each aggregator's activated flexibility to the total revenue of the DSO (the value of flexibility). The MFP of aggregator $h$ is defined as the sum of the dual variables \(\hat{\boldsymbol{\lambda}}_h^\omega\) and \(\check{\boldsymbol{\lambda}}_h^\omega\) associated with constraints \eqref{eq:LinDistFlow:10} and \eqref{eq:LinDistFlow:11} over $\mathcal W$:
\begin{equation}
  \hat{\boldsymbol{\lambda}}_h \triangleq \hat{\boldsymbol{\lambda}}_h^{\text{ru}} + \hat{\boldsymbol{\lambda}}_h^{\text{rd}} \;\;\;\; \text{and} \;\;\;\; \check{\boldsymbol{\lambda}}_h \triangleq \check{\boldsymbol{\lambda}}_h^{\text{ru}} + \check{\boldsymbol{\lambda}}_h^{\text{rd}},
\end{equation}
which reveal the marginal effect of the activated flexibility of aggregator $h$ in reducing the distribution system's net cost. Allocating the total revenue of the DSO according to the MFPs, aggregator $h$ can obtain a payment of:
\begin{equation}
  R_h = \hat{\boldsymbol{\lambda}}_h^{\top} \Delta \hat{\boldsymbol{\phi }}_h + \check{\boldsymbol{\lambda}}_h^{\top} \Delta \check{\boldsymbol{\phi }}_h. \label{eq:settlement}
\end{equation}

The role of MFPs in the flexibility activation process is analogous to that of the distribution locational marginal prices (DLMPs) in the distribution system's retail electricity market. In addition to incentive compatibility, allocating the revenues based on MFPs also has the following property:
\begin{proposition}\label{prop:DSO_non_profit}
  The solution to the DSO optimization model \eqref{eq:market_obj}, \eqref{eq:def_pru_prd}-\eqref{eq:LinDistFlow:14} makes the following inequality hold:
  \begin{equation} 
    {C_{{\text{Base}}}}-{C_{{\text{Energy}}}} + {R_{{\text{Capacity}}}} \ge \sum\nolimits_{h\in \mathcal A} R_h,\label{eq:DSO_non_profit}
  \end{equation}
  where, \(C_{\text{Base}}\) denote the energy cost when the DSO operates on the baseline power profile. The equality holds if all voltage constraints in \eqref{eq:LinDistFlow:6} are not binding.
\end{proposition}

\emph{Proof:} 
The proof can be derived by following the proof in \cite{bobo2021price}, which is trivial and is therefore omitted herein. 

Proposition \ref{prop:DSO_non_profit} ensures that, using the MFPs for revenue allocation, the revenues earned from the TSO are adequate to be allocated to each aggregator (\emph{revenue adequacy}). Besides, in the absence of binding voltage constraints, the revenues obtained from the TSO can be fully allocated to all aggregators in the distribution system, indicating that the DSO does not make any profits in the flexibility activation process. When voltages reach their limits, the DSO will receive a surplus, which is conceptually similar to the TSO's congestion surplus and can be distributed to the aggregators or kept by the DSO for operating the flexibility coordination, upgrading the meters, etc. \cite{mousavi2021dso}.

\vspace{-5pt}
\section{Case Studies}
\subsection{Simulation Setup and Flexibility Cost Coefficients}
Numerical simulations are carried out on a modified IEEE-33 node system \cite{wen2023improved} with $T=24$ and $\Delta_T=1$ hour. There are 32 aggregators distributed at each node except for the root node in the distribution system, each equipped with 20 EVs, 40 HPs, and 1 BESS. Parameters of EVs and BESS are taken from \cite{wen2022aggregate}, while the data of HPs and corresponding buildings come from Z\"urich, Switzerland. The aggregated flexibility model \eqref{eq:agg_flex} is specified as an inner approximated power-energy boundary model \cite{wen2023improved} (with $4T-2$ constraints per aggregator), ensuring that the aggregated power profile can be feasibly disaggregated to each DER.

The compensation for the unsatisfied charging demand of each EV is set to 0.024 EUR/kWh at its departure time and 0.012 EUR/kWh at 23:00. For BESSs, we set two intermediate balancing times of 8:00 and 16:00, at which the compensations for the positive and negative energy imbalances are 0.01 EUR/kWh and 0.02 EUR/kWh, respectively; and we use a hard constraint to ensure that a BESS's energy finally returns to the initial value. For HPs, compensations are set to 0.006\(\cdot \frac{H\Delta_T}{\eta(1-\alpha)}\) EUR/°C for downward temperature adjustment ranges and 0.002\(\cdot \frac{H\Delta_T}{\eta(1-\alpha)}\) EUR/°C for upward ranges in each time slot.\footnote{For the buildings in Z\"urich, this setting generates average compensations of 0.015 EUR/(°C$\cdot$100m$^2$$\cdot$h) for negative temperature adjustment ranges and 0.005 EUR/(°C$\cdot$100m$^2$$\cdot$h) for positive adjustment ranges, respectively. The two numbers mean that for a building with an energy reference area (typically the floor area multiplied by the number of floors) of 100 m$^2$, a negative temperature adjustment range of 1 °C in one hour is compensated with 0.015 EUR, and a positive temperature adjustment range of 1 °C in one hour is compensated with 0.005 EUR.}

\begin{figure}[!t]
  \centering
  \includegraphics[width=3in]{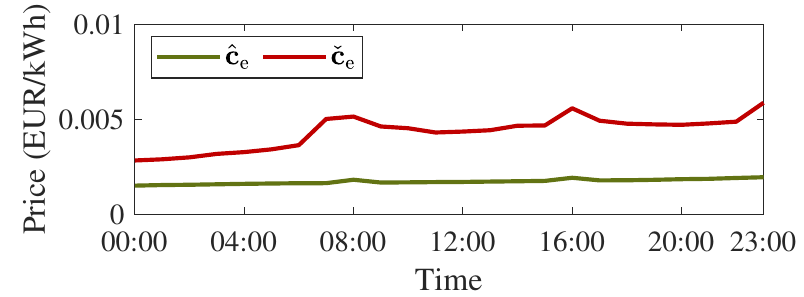}
  \caption{Flexibility cost coefficients of an aggregator.}
  \label{fig:agg_energy_price_coef}
  \vspace{-15pt}
\end{figure}

Based on these settings and the estimation method described in Section III, cost coefficients for the energy adjustment ranges, \(\hat{\mathbf{c}}_\text{e}\) and \(\check{\mathbf{c}}_\text{e}\), of an aggregator are shown in Fig. \ref{fig:agg_energy_price_coef}, while cost coefficients for power adjustment ranges, \(\hat{\mathbf{c}}_\text{p}\) and \(\check{\mathbf{c}}_\text{p}\), are zero. In this figure, \(\check{\mathbf{c}}_\text{e}\) is larger than \(\hat{\mathbf{c}}_\text{e}\), reflecting that energy losses cause more inconvenience to DER users than energy surplus. The observed peaks in \(\check{\mathbf{c}}_\text{e}\), i.e., 8:00, 16:00, and 23:00, are due to compensations for EVs' unsatisfied charging demands and the BESS's energy imbalances.

In the following subsections, we solve the DSO optimization model and analyze the activated flexibility of aggregators, the revenue allocation results, and the impact of the aggregator's flexibility cost on the results. We have considered making a numerical comparison between the proposed methodologies and those in existing studies. However, in the simulation of many existing studies, aggregators are assumed to only report their cost for providing power adjustment range at each time slot to the DSO, without explaining where these cost parameters come from nor reflecting the impact of flexibility activation on the DER devices or/and their users \cite{morstyn2018designing, mousavi2021dso, zhang2022optimal, tsaousoglou2021mechanism, jian2022dlmp}. As analyzed in Section \ref{Sect:flexibility_cost}, power adjustment ranges cannot adequately quantify the flexibility costs associated with time-coupled flexibility regions. Therefore, the DSO optimization models and revenue allocation strategies presented in these works are fundamentally unsuitable for the scenario where the flexibility cost is quantified in a time-coupled manner, making numerical comparisons unviable. The advantage of the proposed flexibility quantification method over traditional power-range-based approaches lies in its consistency with the flexibility models and capability of reflecting the potential inconvenience caused to DER users offering flexibility, which has been comprehensively discussed in Section \ref{Sect:flexibility_cost}. Thus, the numerical simulations here mainly verify the reasonableness of flexibility activation and revenue allocation results calculated by the proposed DSO optimization model and revenue allocation strategy.

\vspace{-10pt}
\subsection{Results of Flexibility Activation and Revenue Allocation}
In the DSO optimization model, the input data of energy and reserve prices at the transmission level are real data from the Netherlands on Jan 2, 2024, where the energy price varied hourly with an average of 63.06 EUR/MWh, and the up- and down-reserve prices were constant and equal to 12.86 EUR/MW and 14.37 EUR/MW, respectively.

\begin{figure}[!t]
  \centering
  \includegraphics[width=2.8in]{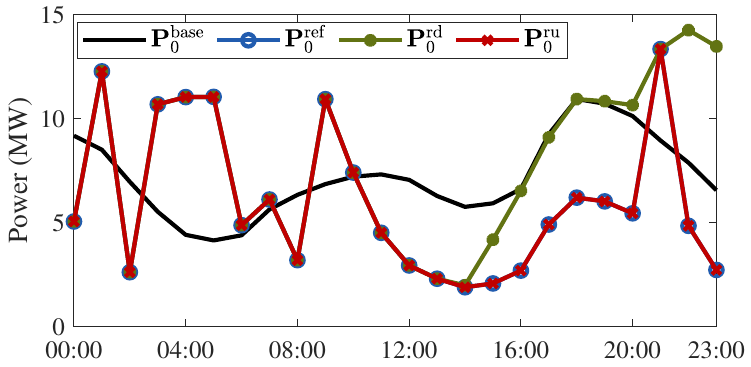}
  \caption{The baseline power profile $\mathbf{P}_0^{\text{base}}$, the reference power profile $\mathbf{P}_0^{\text{ref}}$ and power profiles corresponding to up- and down-reserves, $\mathbf{P}_0^{\text{ru}}$ and $\mathbf{P}_0^{\text{rd}}$, at the root node of the DSO.}
  \label{fig:DSO_ref_rd_ru}
  \vspace{-10pt}
\end{figure}

After solving the DSO optimization model, the DSO's reference power profile and the power profiles corresponding to up- and down-reserve deployments are shown in Fig. \ref{fig:DSO_ref_rd_ru}. The reference power profile $\mathbf{P}_0^{\text{ref}}$ exhibits both upward and downward adjustments from the baseline $\mathbf{P}_0^{\text{base}}$, indicating that the DSO shifts the load for energy arbitrage. The gap between $\mathbf{P}_0^{\text{ru}}$ and $\mathbf{P}_0^{\text{rd}}$ represents reserve capacities. To be more precise, it is the down-reserve (i.e., increasing load), since $\mathbf{P}_0^{\text{ru}}$ coincides exactly with $\mathbf{P}_0^{\text{ref}}$. This coincidence can be explained as follows. Given $\mathbf{P}_0^{\text{ref}} = \mathbf{P}_0^{\text{ru}}$, then changing $\mathbf{P}_0^{\text{ref}}$ to be higher than $\mathbf{P}_0^{\text{ru}}$ (if feasible) will make 1) the flexibility cost remain unchanged, 2) the energy cost increase, 3) the revenue from down-reserve decrease, and 4) the revenue from up-reserve increase. Hence, changing $\mathbf{P}_0^{\text{ref}}$ to be higher than $\mathbf{P}_0^{\text{ru}}$ would not yield a better solution unless the increase in the revenue from up-reserve exceeds the decrease in the revenue from down-reserve plus the increase in energy cost. This condition, however, is not met in the tested case because the energy prices are significantly higher than the up-reserve prices.

\begin{figure}[t!]
  \begin{centering}
  \subfigure[Power adjustments]{\includegraphics[width=1.65in]{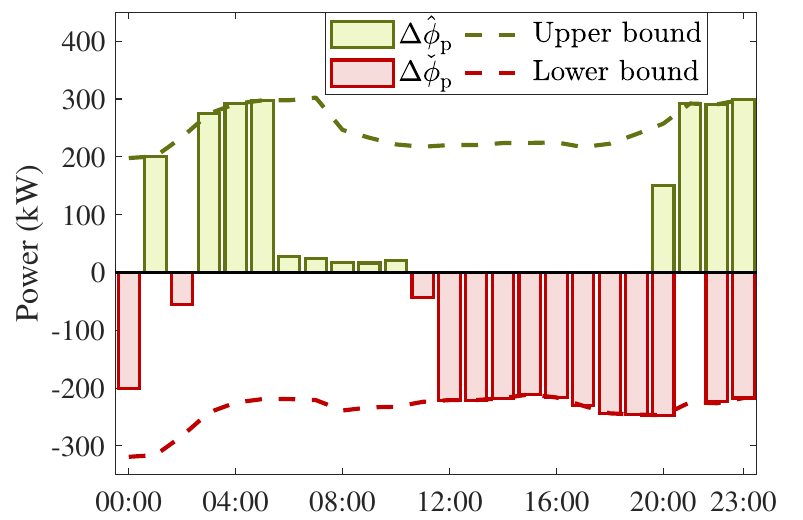}\label{fig:result_agg:power_deviations}}
  \subfigure[Energy adjustments]{\includegraphics[width=1.65in]{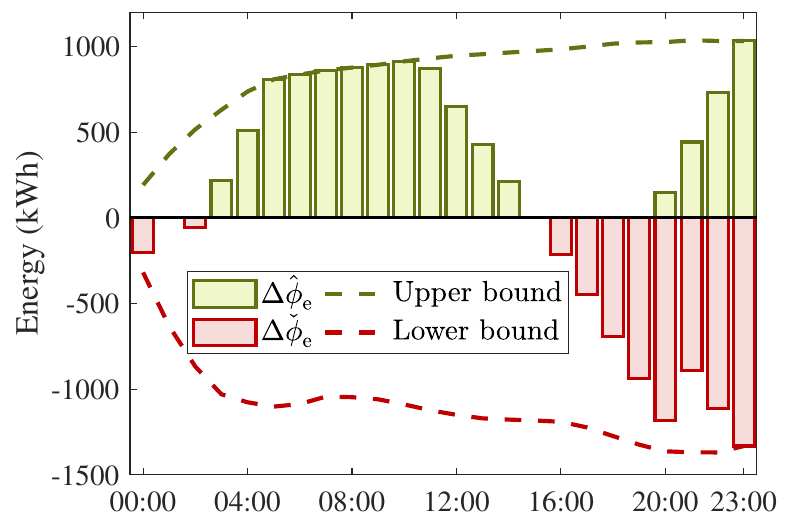}\label{fig:result_agg:energy_deviations}}\\
  \subfigure[MFPs for power]{\includegraphics[width=1.65in]{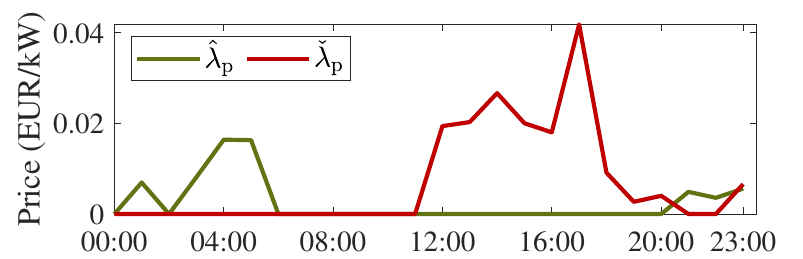}\label{fig:result_agg:power_MFPs}}
  \subfigure[MFPs for energy]{\includegraphics[width=1.65in]{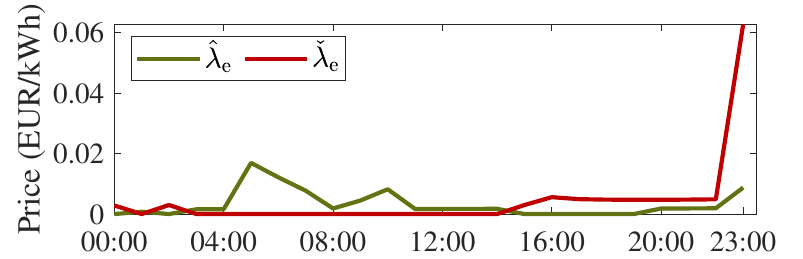}\label{fig:result_agg:energy_MFPs}}\\
  \end{centering}
  \caption{The activated flexibility and MFPs. $\Delta \hat{\boldsymbol{\phi}}_{\text{p}}$ and $\Delta \check{\boldsymbol{\phi}}_{\text{p}}$: power parts of the flexibility trajectory's upward and downward adjustment ranges; $\Delta \hat{\boldsymbol{\phi}}_{\text{e}}$ and $\Delta \check{\boldsymbol{\phi}}_{\text{e}}$: accumulated energy consumption parts of the flexibility trajectory's upward and downward adjustment ranges; $\hat{\boldsymbol{\lambda}}_{\text{p}}$ and $ \check{\boldsymbol{\lambda}}_{\text{p}}$: MFPs corresponding to the upward and downward power adjustment ranges; $\hat{\boldsymbol{\lambda}}_{\text{e}}$ and $ \check{\boldsymbol{\lambda}}_{\text{e}}$: MFPs corresponding to the upward and downward adjustment ranges in accumulated energy consumption.}
  \label{fig:result_agg}
  \vspace{-15pt}
\end{figure}

We take the aggregator at node 2 as an example to illustrate the activated flexibility and the corresponding MFPs (see Fig. \ref{fig:result_agg}). The pattern of flexibility activation in Fig. \ref{fig:result_agg}, i.e., times and directions of the adjustments, is consistent with the adjustment pattern of power profiles at the root node from the baseline in Fig. \ref{fig:DSO_ref_rd_ru}. For instance, at 20:00, 22:00, and 23:00, $\mathbf{P}_0^{\text{ru}}$ and $\mathbf{P}_0^{\text{rd}}$ in Fig. 6 exhibit different adjustment directions relative to the baseline, which is reflected in Fig. \ref{fig:result_agg:power_deviations}, where both the upward and downward power adjustments at these times are nonzero, also correlating with the nonzero energy adjustments in Fig. \ref{fig:result_agg:energy_deviations} after 20:00. The power and energy parts of the aggregator's MFPs are shown in Figs. \ref{fig:result_agg:power_MFPs} and \ref{fig:result_agg:energy_MFPs}. While the power part in the aggregator's flexibility cost coefficients are all zero, the power part in MFPs is nonzero because power adjustments contribute to the DSO's total revenue.

We next discuss the revenue allocation results based on MFPs. First, we study the case without taking the voltage constraints into account, i.e. removing \eqref{eq:LinDistFlow:6} from the problem formulation. In this case, the total revenue of the DSO from energy arbitrage and reserve capacity provision in the transmission system, i.e., the left-hand side of inequality \eqref{eq:DSO_non_profit}, is 4410.98 EUR. Distributing the revenues to each aggregator according to \eqref{eq:settlement}, the sum of each aggregator's payments $R_h$, i.e., the right-hand side of inequality \eqref{eq:DSO_non_profit}, is also 4410.98 EUR, exactly the same as the total revenue of the DSO. This result confirms that the proposed MFPs for settlement ensure the DSO's balance of payments when the voltage constraints are not binding (equivalent to being absent). However, if we use only the dual variables corresponding to power range constraints for revenue allocation, the total payment allocated to all aggregators is only 1118.46 EUR. This result indicates that the primary factor affecting the total revenue of the distribution system is not the power adjustment range but the energy adjustment range in this simulation. Therefore, it is necessary to use MFPs, which include both the power and energy parts, for revenue allocation.

Then, by adding back the voltage constraints and studying a case where one of these constraints becomes binding, the DSO's total revenue and the aggregators' total payments are 4410.98 EUR and 4406.93 EUR, respectively, generating a surplus of 4.05 EUR. The surplus has been discussed right after Proposition \ref{prop:DSO_non_profit}.
\vspace{-15pt}
\subsection{Impact of the Aggregators' Flexibility Costs on the Results}
\begin{figure}[!t]
  \centering
  \includegraphics[width=3in]{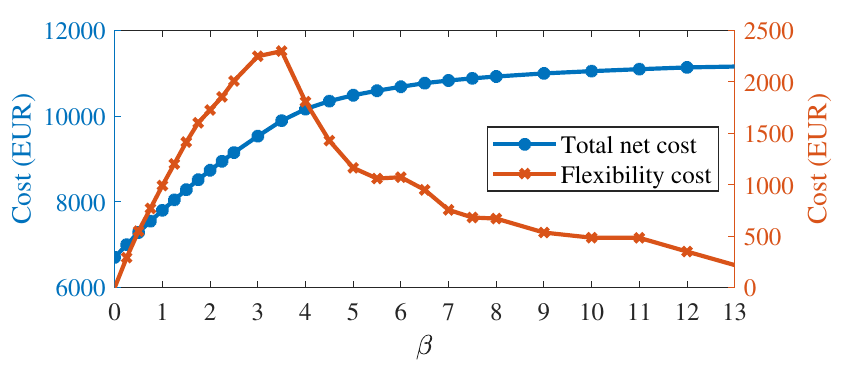}
  \caption{Changes in costs of the DSO with changes in the flexibility cost coefficients.}
  \label{fig:res_changing_bid_together}
  \vspace{-15pt}
\end{figure}

This section analyzes the impact of the aggregators' flexibility costs on the DSO's flexibility activation results. First, the case where the flexibility costs of all aggregators change is tested. We multiply the cost coefficients of all aggregators by a parameter $\beta$ and analyze how the calculated total net cost and flexibility cost of the DSO change with $\beta$. The results, as shown in Fig. \ref{fig:res_changing_bid_together}, indicate that as the flexibility cost coefficients increase, i.e., $\beta$ increases, the DSO's total net cost also increases as expected. Second, the DSO's overall flexibility cost initially increases and then decreases as the flexibility cost coefficients increase. This result is because the total flexibility cost is the product of the cost coefficients and the activated flexibility, and the latter decreases as the flexibility cost coefficients increase.

\begin{figure}[!t]
  \centering
  \includegraphics[width=2.8in]{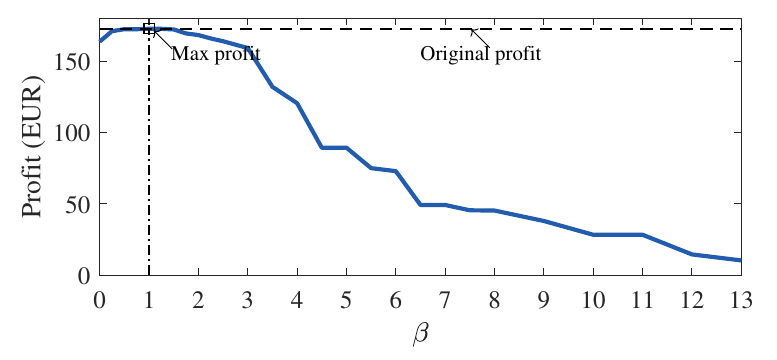}
  \caption{Changes in the profit of the aggregator at node 2 with its reported flexibility cost.}
  \label{fig:res_changing_bid_alone}
  \vspace{-15pt}
\end{figure}

Next, we test the case where only the reported cost coefficients of one aggregator change. It is assumed that the aggregators' original cost coefficients are their true cost coefficients. Keeping the cost coefficients of other aggregators constant, we change the reported cost coefficients of the aggregator at node 2 by multiplying it with the parameter $\beta$, implying that this aggregator does not report its true cost coefficients. After the DSO optimization problem is solved, the net profit of the aggregator (the received payment of the aggregator $R_h$ minus the aggregator's true flexibility cost) can be obtained, as shown in Fig. \ref{fig:res_changing_bid_alone}. It is observed that the maximum net profit for the aggregator occurs at $\beta=1$, i.e., when the aggregator reports its true flexibility costs. This result verifies that the proposed flexibility activation model can encourage aggregators to report their true cost coefficients.
\vspace{-10pt}
  \section{Conclusions}
   This paper proposes a quantification method for the costs of the time-coupled flexibilities from DERs. For individual DERs, we propose to compute their flexibility cost based on adjustment ranges in both power and accumulated energy consumption. From the aggregator's standpoint, we propose to compute their flexibility cost based on adjustment ranges in the flexibility trajectory. The proposed quantification method is unified for individual and aggregated DERs. It can also capture the time-coupled characteristics of the flexibility regions, which are not reflected in the traditional power-range-based approaches.
  
  A DSO optimization model is then designed to activate the flexibility to participate in the transmission system operation for energy arbitrage and reserve capacity provision. From this optimization model, we derive the concept of MFPs, which quantify the marginal contributions of the activated flexibility to the DSO's total revenue in the TSO-DSO coordination (the value of flexibility). The MFPs are then used to allocate the revenues obtained from the TSO to each aggregator in the distribution system. The proposed DSO optimization model and the revenue allocation strategy can ensure incentive compatibility, revenue adequacy, and a non-profit DSO.
  
  A potential limitation of this work lies in the calculation of the aggregators' cost coefficients, which are estimated using a weighted average method in this paper and may be improved in future works. Future work can also focus on real-time disaggregation and control to better exploit flexibility in TSO-DSO coordination.
  \appendix[Proof of Proposition 1]
\setcounter{equation}{0}
\renewcommand\theequation{A.\arabic{equation}}
We first reformulate the temperature evolution model \eqref{eq:temp_evolution} using a time-sequential form, as follows:
{\small
  \begin{align}
    {\theta _T} - \alpha {\theta _{T - 1}} &= (1 - \alpha )(\theta _T^{{\text{amb}}} + \frac{\eta }{H} \cdot {p_T}),\nonumber\\
    {\theta _{T - 1}} - \alpha {\theta _{T - 2}} &= (1 - \alpha )(\theta _{T - 1}^{{\text{amb}}} + \frac{\eta }{H} \cdot {p_{T - 1}}),\nonumber\\
     &\cdots \nonumber\\
    {\theta _2} - \alpha {\theta _1} &= (1 - \alpha )(\theta _2^{{\text{amb}}} + \frac{\eta }{H} \cdot {p_2}),\nonumber\\
    {\theta _1} - \alpha {\theta _0} &= (1 - \alpha )(\theta _1^{{\text{amb}}} + \frac{\eta }{H} \cdot {p_1}),\nonumber
\end{align}
}where $\theta_0$ is the initial indoor temperature. These equations derive the relationship between the indoor temperature in the \(t\)-th time slot and the preceding power sequence \(p_1, p_2, \cdots, p_t\):
\begin{equation}
  \small
  {\theta _t} = (1 - \alpha )\sum\limits_{\tau  = 1}^t {{\alpha ^{t - \tau }}(\theta _\tau ^{{\text{amb}}} + \frac{\eta }{H} \cdot {p_\tau })}  + {\alpha ^t}{\theta _0},\forall t\in [T]. \label{eq:temp_from_p_t}
\end{equation}
Writing \eqref{eq:temp_from_p_t} in matrix form gives:
\begin{equation}
  \boldsymbol{\theta}  = (1 - \alpha ){\mathbf{A}}{\boldsymbol{\theta} ^{{\text{amb}}}} + (1 - \alpha )\frac{\eta }{H} \cdot {\mathbf{Ap}} + {\boldsymbol{\alpha }}{\theta _0}, \label{eq:temp_from_p_matrix}
\end{equation}
where
\begin{equation}
  \small
  {\mathbf{A}} \triangleq \left[ {\begin{array}{cccc}
    1&{}&{}&{}\\
    \alpha &1&{}&{}\\
     \vdots & \ddots & \ddots &{}\\
    {{\alpha ^{T - 1}}}& \cdots &\alpha &1
    \end{array}} \right],{\boldsymbol{\alpha }} \triangleq \left[ {\begin{array}{c}
    \alpha \\
    {{\alpha ^2}}\\
     \vdots \\
    {{\alpha ^T}}
    \end{array}} \right]. \nonumber
\end{equation}
Equation \eqref{eq:temp_from_p_matrix} holds for the baseline power profile $\mathbf{p}^{\text{base}}$ and the temperature set point $\boldsymbol{\theta}^{\text{set}}$, that is:
\begin{equation}
  \boldsymbol{\theta}^{\text{set}}  = (1 - \alpha ){\mathbf{A}}{\boldsymbol{\theta} ^{{\text{amb}}}} + (1 - \alpha )\frac{\eta }{H} \cdot {\mathbf{Ap}^{\text{base}}} + {\boldsymbol{\alpha }}{\theta _0}. \label{eq:temp_from_p_matrix_base}
\end{equation}
Subtracting \eqref{eq:temp_from_p_matrix_base} from \eqref{eq:temp_from_p_matrix} yields:
\begin{equation}
  \boldsymbol{\theta}-\boldsymbol{\theta}^{\text{set}}  = (1 - \alpha )\frac{\eta }{H} \cdot {\mathbf{A}(\mathbf{p}-\mathbf{p}^{\text{base}})}. \label{eq:temp_from_p_matrix_deviation}
\end{equation}
As the matrix $\mathbf A$ is invertible, namely its inverse is:
\begin{equation}
  \small
  {{\mathbf{A}}^{ - 1}} = \left[ {\begin{array}{cccc}
    1&{}&{}&{}\\
    { - \alpha }&1&{}&{}\\
    {}& \ddots & \ddots &{}\\
    {}&{}&{ - \alpha }&1
    \end{array}} \right],\nonumber
\end{equation}
expression \eqref{eq:temp_from_p_matrix_deviation} can be transformed to:
\begin{equation}
  {\mathbf{p}} - \mathbf{p}^{\text{base}} = \frac{H}{\eta (1 - \alpha )}{{\mathbf{A}}^{ - 1}}\left( {{\boldsymbol{\theta }} -\boldsymbol{\theta}^{\text{set}} } \right). \label{eq:p_from_temp_matrix_deviation}
\end{equation}

We now consider the temperature adjustment limits:
\begin{equation}
  -\Delta \check{\boldsymbol{\theta}} \le \boldsymbol{\theta}-\boldsymbol{\theta}^{\text{set}} \le \Delta \hat{\boldsymbol{\theta}}. \label{eq:temp_range_vector}
\end{equation}
We first prove the second half of Proposition 1 using a contradiction. Suppose there exists an \( \mathbf F \in \mathbb R ^{T\times T}\) that satisfies the conditions. We set \(\Delta \hat{\boldsymbol{\theta}} = \mathbf 0\) and \(\Delta \check{\boldsymbol{\theta}} > \mathbf 0\), so that the power profile $\mathbf{p}$ should satisfy \(-\mathbf F \Delta \check{\boldsymbol{\theta}} \leq \mathbf{p}-\mathbf{p}^{\text{base}} \leq \mathbf 0\). Let $\theta_1  = \theta_1^{\text{set}} - \Delta \check{\theta}_1$ and $\theta_2  = \theta_2^{\text{set}}$, then the first two rows of Equation \eqref{eq:p_from_temp_matrix_deviation} becomes:
\begin{align}
  \left[ {\begin{array}{cc}
    {{p_1} - p_1^{{\text{base}}}}\\
    {{p_2} - p_2^{{\text{base}}}}
    \end{array}} \right] &= \frac{H}{{\eta (1 - \alpha )}}\left[ {\begin{array}{cc}
    1&{}\\
    { - \alpha }&1
    \end{array}} \right]\left[ {\begin{array}{cc}
    { - \Delta {\theta _1}}\\
    0
    \end{array}} \right] \nonumber\\
    &= \frac{H}{{\eta (1 - \alpha )}}\left[ {\begin{array}{cc}
    { - \Delta {\check{\theta} _1}}\\
    {\alpha \Delta {\check{\theta} _1}}
    \end{array}} \right]. \nonumber
\end{align}
Thus, ${{p_2} - p_2^{{\text{base}}}} = \frac{H \alpha \Delta {\check{\theta} _1} }{{\eta (1 - \alpha )}}>0$, which contradicts \(\mathbf{p}-\mathbf{p}^{\text{base}} \leq \mathbf 0\), proving the second half of Proposition 1. 

Readers might wonder why we cannot multiply both sides of \eqref{eq:temp_range_vector} by \(\frac{H}{\eta (1 - \alpha )}{{\mathbf{A}}^{ - 1}}\) to generate a constraint on \(\mathbf{p}-\mathbf{p}^{\text{base}}\). This is because the elements of \({\mathbf{A}}^{-1}\) are not all positive, and thus, multiplying both sides of \eqref{eq:temp_range_vector} by \(\frac{H}{\eta (1 - \alpha )}{{\mathbf{A}}^{ - 1}}\) does not necessarily preserve the inequality's direction. Nevertheless, we will show that this operation becomes feasible when transforming the temperature adjustment ranges $\Delta \hat{\boldsymbol{\theta}}$ and $\Delta \check{\boldsymbol{\theta}}$ into the energy adjustment ranges $\Delta \hat{\mathbf{e}}$ and $\Delta \check{\mathbf{e}}$, and it will prove the first half of Proposition 1. To this end, we write the matrix form of the energy-power relationship \eqref{eq:power_energy_relationship}:
\begin{equation}
  \mathbf e = \Delta_T \mathbf {Lp}, \mathbf e^{\text{base}} = \Delta_T \mathbf {Lp}^{\text{base}},  \label{eq:power_energy_relationship_matrix}
\end{equation}
where \(L\) is a $T\times T$ lower triangular matrix with all lower triangle elements equal to one.
Multiplying both sides of \eqref{eq:p_from_temp_matrix_deviation} by $\Delta_T \mathbf {L}$ gives
\begin{equation}
  {\mathbf{e}} - \mathbf{e}^{\text{base}} = \frac{\Delta_T H}{\eta(1 - \alpha ) }{{\mathbf{LA}}^{ - 1}}\left( {{\boldsymbol{\theta }} -\boldsymbol{\theta}^{\text{set}} } \right), \label{eq:e_from_temp_matrix_deviation}
\end{equation}
where 
\begin{equation}
  \small
  {\mathbf{L}}{{\mathbf{A}}^{ - 1}} = \left[ {\begin{array}{cccc}
    1&{}&{}&{}\\
    {1 - \alpha }&1&{}&{}\\
     \vdots & \ddots & \ddots &{}\\
    {1 - \alpha }& \cdots &{1 - \alpha }&1
    \end{array}} \right].\nonumber
\end{equation}
All the components of ${\mathbf{L}}{{\mathbf{A}}^{ - 1}} $ are non-negative since $0<a<1$ by definition, hence The direction of inequality \eqref{eq:temp_range_vector} is preserved when multiplying both sides by $\frac{\Delta_T H}{\eta(1 - \alpha ) }{\mathbf{L}}{{\mathbf{A}}^{ - 1}} $, i.e.,
\begin{equation}
  -\frac{\Delta_T H}{\eta(1 - \alpha ) }{\mathbf{L}}{{\mathbf{A}}^{ - 1}} \Delta \hat{\boldsymbol{\theta}} \le {\mathbf{e}} - \mathbf{e}^{\text{base}} \le \frac{\Delta_T H}{\eta(1 - \alpha ) }{\mathbf{L}}{{\mathbf{A}}^{ - 1}} \Delta \check{\boldsymbol{\theta}}. \label{eq:e_range_vector}
\end{equation}Let $\mathbf D \triangleq \frac{\Delta_T H}{\eta(1 - \alpha ) }{\mathbf{L}}{{\mathbf{A}}^{ - 1}} $, then $\Delta \hat{\mathbf{e}} = {\mathbf{D}}\Delta \hat{\boldsymbol{\theta}}$ and $\Delta \check{\mathbf{e}} = {\mathbf{D}}\Delta \check{\boldsymbol{\theta}}$, which complete the proof of Proposition 1.

\small
\bibliographystyle{IEEEtran}

\bibliography{ref}


\end{document}